\documentclass[sigconf, nonacm]{acmart}

\AtBeginDocument{%
  }

\setcopyright{acmlicensed}
\copyrightyear{2026}
\acmYear{2026}
\acmDOI{XXXXXXX.XXXXXXX}
\acmConference[GRAIL '26]{Workshop on Generative, Retrieval-augmented, and Agentic Intelligence for Personalization}{November 2026}{Rome, Italy}
\acmISBN{978-1-4503-XXXX-X/18/06}

\usepackage{amsmath}    
\usepackage{booktabs}   
\usepackage{multirow}   
\usepackage{graphicx}
\usepackage{array}
\usepackage{float}

\begin{document}

\title{Routing Between Generative and Collaborative User Profiles:\\
A Serving-Time Gate for Controllable Novelty}

\author{Milad Sabouri}
\orcid{0009-0009-8219-5596}
\affiliation{%
   \institution{DePaul University / Comcast Technology AI}
   \city{Chicago}
   \state{IL}
   \country{USA}}
\email{msabouri@depaul.edu}
\email{milad_sabouri@comcast.com}

\author{Neeraj Sharma}
\orcid{0009-0008-7940-1523}
\affiliation{%
   \institution{Comcast Technology AI}
   \city{Sunnyvale}
   \state{CA}
   \country{USA}}
\email{neeraj_sharma@comcast.com}

\author{Sardar Hamidian}
\orcid{0009-0007-5377-0374}
\affiliation{%
   \institution{Comcast Technology AI}
   \city{Washington}
   \state{DC}
   \country{USA}}
\email{sardar_hamidian@comcast.com}

\author{Shaghayegh Agah}
\authornote{Corresponding author.}
\orcid{0000-0002-7063-9251}
\affiliation{%
   \institution{Comcast Technology AI}
   \city{Sunnyvale}
   \state{CA}
   \country{USA}}
\email{shaghayegh_agah@comcast.com}

\renewcommand{\shortauthors}{Sabouri et al.}

\begin{abstract}
Large language models (LLMs) enable rich semantic user profiles for recommendation, but such profiles are more expensive to generate and are not necessarily desirable to deploy uniformly. We study whether LLM-generated profiles can instead be invoked selectively within a production recommendation pipeline. Using a real-world streaming dataset covering movies, TV shows, and sports content, we train a serving-time routing gate that assigns each user to either a collaborative sequential recommendation model or a recommendation model driven by an LLM-generated profile. The gate uses only serving-time features and learns to identify users for whom profile-based routing can increase Novelty@10 while preserving ranking relevance. A routing threshold controls how aggressively users are sent to the generative model, exposing a tunable novelty--relevance trade-off. At an overall NDCG-loss budget of 5\%, the learned gate increases Novelty@10 by 6.5\% while routing 12.5\% of users, outperforming simple heuristic and random routing policies at comparable relevance cost. These results show that LLM-generated user profiles can serve as a controllable complement to collaborative recommendation, while results with non-generative semantic profiles indicate that the benefit stems from selective routing rather than LLM generation alone.

\end{abstract}



\maketitle

\section{Introduction}
\label{sec:introduction}

Collaborative sequential recommenders are a strong foundation for modern recommender systems because they learn regularities in users' interaction sequences and typically provide highly relevant recommendations. Their reliance on observed interaction patterns, however, can favor well-covered and popular content. LLM-generated user profiles provide a complementary representation: an LLM summarizes a user's interaction history into a natural-language description of their preferences, which is then encoded into a semantic space for recommendation. Prior work has shown that such profiles can capture high-level and temporally varying user preferences~\cite{sabouri2025effectiveness}. More recent production-scale evaluation found that their effectiveness relative to aggregate semantic profiles depends on user behavior and can involve substantial beyond-accuracy trade-offs, motivating adaptive rather than uniform use~\cite{sabouri2026llm}. In our setting, profile-based recommendation models substantially increase exposure to less-popular content, but uniform deployment incurs a large loss in ranking relevance. LLM-based profiling also introduces additional generation and refresh cost, further motivating selective rather than global use.

This tension motivates a different role for generative user profiles. Rather than treating them as a replacement for collaborative recommendation, we ask whether they can be used \emph{selectively}. A system that can identify users for whom profile-based routing is likely to provide additional novelty can retain the collaborative model as the default and
invoke a recommendation model driven by an LLM-generated profile only when the expected trade-off is favorable. This is particularly relevant in production settings, where objectives extend beyond relevance alone and operators may be willing to exchange a bounded amount of relevance for increased exposure to less-popular content.

We study this problem on a real-world production streaming dataset covering movies, TV shows, and sports content. We formulate routing as a per-user prediction problem between a collaborative sequential recommendation model and candidate profile-based recommendation models, including the LLM-generated Narrative and TemporalNarrative representations and a
non-generative Centroid profile. The routing gate observes only serving-time information derived from the user's historical behavior and learned representations. It is trained to identify users for whom routing to a candidate model can increase recommendation novelty while preserving relevance. At inference time, the gate produces a routing score, and varying its decision threshold controls how aggressively users are routed away from the collaborative model. The result is a controllable novelty--relevance frontier rather than a single fixed hybrid policy.

Our experiments show that selective routing can improve novelty while bounding overall relevance degradation. Using Narrative as the primary generative case study, a policy operating within an approximately 5\% NDCG-loss budget increases Novelty@10 by 6.5\% while routing only 12.5\% of users. At comparable relevance cost, the learned gate also produces
larger novelty gains than random routing and simple heuristics. These results suggest that the value of generative profiling depends not only on how the representation is constructed, but also on deciding \emph{when} to use it.

Our contributions are twofold: (1) we formulate selective deployment of profile-based recommendation models, including those driven by LLM-generated profiles, as a serving-time routing problem with novelty as the reward and ranking relevance as a guardrail; and (2) we demonstrate on real-world production streaming data that a learned gate exposes a tunable novelty--relevance trade-off and allocates a fixed relevance budget more effectively than heuristic or random routing.

\section{Related Work}
\label{sec:related_work}

\paragraph{Semantic and generative user profiling.}
Content-based recommenders commonly construct user profiles by aggregating item representations~\cite{lops2011content}, with sentence encoders such as Sentence-BERT providing dense semantic representations~\cite{reimers2019sbert}. More recent approaches use LLMs to generate natural-language summaries of user preferences that can be embedded for recommendation and explanation~\cite{lubos2024llmexpl,ma2024xrec,sabouri2025exum}. Temporal extensions generate separate short- and long-term preference summaries and combine their representations through learned attention~\cite{sabouri2025effectiveness}. Production-scale evaluation has further shown that the relative effectiveness of LLM-generated and aggregate semantic profiles can vary across user consumption regimes and beyond-accuracy objectives~\cite{sabouri2026llm}. We build on these profiling strategies but study a different question: when a generative representation should be served instead of a collaborative sequential representation.

\paragraph{Beyond-accuracy objectives.}
Recommendation quality extends beyond ranking relevance to properties such as novelty and diversity~\cite{vargas2011rank,castells2015novelty}, while popularity bias remains a persistent concern in recommender systems~\cite{abdollahpouri2019popularity}. Rather than optimizing novelty globally, we treat Novelty@10 as a routing reward while constraining degradation in NDCG@10, yielding operating points that explicitly expose the relevance--novelty trade-off.
\paragraph{Adaptive routing and selective computation.}
Learned gating has been used to adaptively combine specialized experts in recommendation and multi-task learning~\cite{ma2018modeling}, while model-cascade approaches selectively invoke models to balance predictive quality and computational cost~\cite{chen2023frugalgpt}. Our setting differs in that the gate makes a serving-time, per-user choice between collaborative and profile-based recommendation models and varies its routing threshold to operate under an explicit relevance budget.

\section{Methodology}
\label{sec:methodology}

We learn a per-user routing policy between the default collaborative sequential recommendation model, denoted by $C$, and a candidate semantic recommendation model, denoted by $G$. The semantic model uses one of the user representations described below. The router observes only information available at serving time and selects which recommendation model produces
the final recommendation list for each user.

\subsection{User Representations}
\label{sec:representations}

We consider two LLM-generated user profiles following prior temporal profiling work~\cite{sabouri2025effectiveness} and its subsequent production-scale evaluation~\cite{sabouri2026llm}. Narrative generates a single natural-language summary from the user's interaction history and encodes it with SBERT. TemporalNarrative generates separate short- and long-term preference summaries, encodes them with SBERT, and combines the
resulting representations through learned attention. We additionally consider Centroid, a non-generative semantic profile obtained by mean-pooling the embeddings of items in the user's interaction history. Each representation defines a separate semantic recommendation model, and a separate routing gate is trained for each one.

\subsection{Learned Routing Gate}
\label{sec:routing_gate}

For each user $u$, the gate observes a serving-safe feature vector $\mathbf{x}_u$ consisting of history length and its logarithm, mean and median popularity of consumed items, the fraction of below-median-popularity items (niche ratio), and the norms of the collaborative and candidate semantic user representations.

The gate is trained to identify users for whom routing from $C$ to $G$ increases recommendation novelty without reducing ranking relevance. We first compute the per-user differences
\begin{equation}
\Delta \mathrm{Nov}_u =
\mathrm{Novelty}^{G}_u-\mathrm{Novelty}^{C}_u,\quad
\Delta \mathrm{NDCG}_u =
\mathrm{NDCG}^{G}_u-\mathrm{NDCG}^{C}_u .
\end{equation}

The binary routing target is then defined as
\begin{equation}
y_u =
\mathbb{I}
\left[
\Delta \mathrm{Nov}_u > 0
\land
\Delta \mathrm{NDCG}_u \geq 0
\right],
\end{equation}
where $\mathbb{I}[\cdot]$ equals one when both conditions are satisfied and zero otherwise.

A Gradient Boosting classifier trained with binary log-loss estimates $q_u=P(y_u=1\mid\mathbf{x}_u)$. Given a routing threshold $\tau$, the serving policy is
\begin{equation}
\pi_\tau(u)=
\begin{cases}
G, & q_u \geq \tau,\\
C, & q_u < \tau.
\end{cases}
\end{equation}

Varying $\tau$ controls how many users are routed to the semantic model and consequently the novelty--relevance trade-off. We sweep $\tau$ and report operating points under different overall NDCG-loss budgets relative to always serving the collaborative model. These budgets are used only to select operating points after training; they are not part of the classifier objective. All routing predictions are generated using five-fold out-of-fold evaluation, so a user's observed recommendation outcomes are not used to train the gate that produces that user's routing score.
\section{Experimental Evaluation}
\label{sec:experiments}

\subsection{Experimental Setup}
\label{sec:experimental_setup}

\begin{figure*}[t]
    \centering
    \begin{minipage}[t]{0.48\textwidth}
        \centering
        \textbf{(a)}\\
        \includegraphics[width=\linewidth]{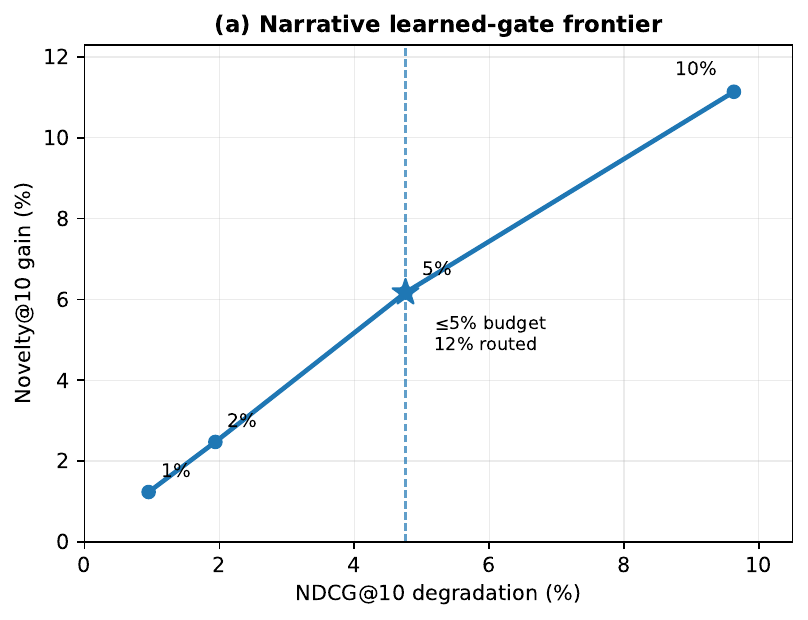}
    \end{minipage}
    \hfill
    \begin{minipage}[t]{0.48\textwidth}
        \centering
        \textbf{(b)}\\
        \includegraphics[width=\linewidth]{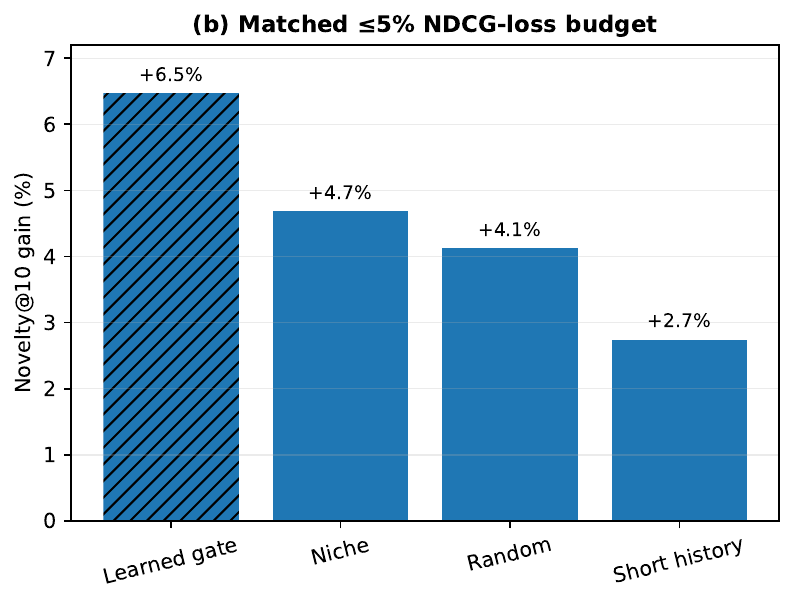}
    \end{minipage}
    \caption{Controllable novelty--relevance routing with the Narrative profile at $K=10$. (a) Sweeping the learned-gate probability threshold produces a controllable frontier between NDCG degradation and Novelty gain. (b) Using a finer coverage-ranked sweep, each policy selects its highest-Novelty feasible operating point under the same overall NDCG-loss budget of at most 5\%; the learned gate obtains the largest Novelty gain.}
    \label{fig:routing_tradeoff}
\end{figure*}

\textbf{Data and recommendation setting.}
We evaluate on a sample of 10,000 users from a real-world production streaming platform covering movies, TV shows, and sports content. The default recommendation model is a collaborative sequential model that predicts a user's next interaction from historical behavior. The candidate profile-based models use Centroid, Narrative, or TemporalNarrative as described in Section~\ref{sec:representations}. All three use a common semantic item encoder, and recommendation is performed over the full eligible catalog. Representation-generation details are provided in Appendix~\ref{app:routing_details}. Because routing labels require outcomes from both the collaborative and candidate models, all policy comparisons are performed on the
common test population for which outcomes are available from the collaborative model and all three profile-based models. This ensures that every routing policy is evaluated on the same users. The recommendation models and user representations are held fixed throughout the routing experiment; only the routing gate is learned.

\textbf{Metrics and reporting.}
We measure ranking relevance using NDCG@10 and novelty using Novelty@10, defined as the mean self-information of the recommended items,
\[
\mathrm{Novelty@}K(u)=
\frac{1}{K}\sum_{i\in R_u^K}-\log_2 p(i),
\]
where $p(i)$ is item popularity estimated from training interactions~\cite{vargas2011rank}. Higher Novelty therefore indicates
exposure to less-popular items. We focus on Novelty rather than intra-list diversity because the collaborative and profile-based
recommendation models operate in different representation geometries, making diversity less directly comparable across them. To preserve production confidentiality, all recommendation-performance results are reported as relative changes with respect to always serving the collaborative model.

\textbf{Routing evaluation.}
A separate Gradient Boosting gate is trained for each candidate profile. Routing scores are generated using five-fold out-of-fold evaluation, such that each user's score is produced by a gate trained on outcome-derived labels from the other folds. The held-out user's observed recommendation outcomes are therefore not used to train the gate that produces that user's routing score.

We compare the learned Narrative gate against three simple policies: a \textit{niche} heuristic that prioritizes users with higher niche ratios, a \textit{short-history} heuristic that prioritizes users with shorter interaction histories, and uniform random routing. The two heuristic policies also test whether routing on a single intuitive user signal can reproduce the benefit of the multivariate gate. Because different policies incur different relevance costs per routed user, we compare them at matched overall NDCG loss rather than matched routing rates. At the 5\% budget, each policy is allowed to select the operating point on its frontier that yields the greatest Novelty gain while remaining within the relevance constraint. Bootstrap 95\% confidence intervals are reported for the Narrative operating points, and paired Wilcoxon signed-rank tests~\cite{wilcoxon1945individual} are applied to routed-user novelty differences. Additional implementation details are provided in Appendix~\ref{app:routing_details}, and routing results for Centroid and TemporalNarrative are reported in Appendix~\ref{app:additional_routing}.

\subsection{Results and Discussion}
\label{sec:results}

\textbf{Uniform profile-based deployment motivates selective routing.}
We first examine the trade-off obtained by serving each profile-based recommendation model uniformly. Table~\ref{tab:semantic_arms} shows that all three substantially increase Novelty@10, but at a large cost in NDCG@10.

\begin{table}[t]
\centering
\small
\renewcommand{\arraystretch}{1.03}
\caption{Semantic-arm characteristics at $K=10$. Uniform-deployment columns serve the indicated semantic representation to all users and report changes relative to always-collaborative. Routing signal reports the out-of-fold ROC-AUC improvement above chance for the learned gate.}
\label{tab:semantic_arms}
\begin{tabular*}{\columnwidth}{@{\extracolsep{\fill}}lccc@{}}
\toprule
& \multicolumn{2}{c}{Uniform deployment}
& \multicolumn{1}{c}{Learned gate} \\
\cmidrule(lr){2-3}\cmidrule(l){4-4}
Semantic arm
& $\Delta$NDCG
& $\Delta$Novelty
& AUC lift \\
\midrule
\multicolumn{4}{@{}l}{\textit{Non-generative}} \\
\hspace{0.8em}Centroid
& $-85.1\%$
& $+69.0\%$
& +15.9\,pp \\

\cmidrule(lr){1-4}

\multicolumn{4}{@{}l}{\textit{Generative}} \\
\hspace{0.8em}TemporalNarrative
& $-87.5\%$
& $+74.7\%$
& +15.9\,pp \\

\hspace{0.8em}Narrative
& $-94.7\%$
& $+75.7\%$
& +18.4\,pp \\
\bottomrule
\end{tabular*}
\end{table}

Uniform deployment confirms that both generative and non-generative semantic profiles substantially increase novelty but incur large relevance losses. The LLM-generated profiles yield the largest uniform novelty gains, although at greater relevance cost. The routing AUCs indicate that serving-time features contain predictive signal for each candidate model; because each gate predicts an arm-specific outcome, we do not use cross-arm AUC differences to rank the candidate models. We use Narrative as the primary case study because our focus is the selective deployment of an LLM-generated profile, while Appendix~\ref{app:additional_routing} reports corresponding results for Centroid and TemporalNarrative.

\textbf{The learned gate exposes a controllable novelty--relevance trade-off.}
Rather than serving Narrative uniformly, we vary the routing threshold to control how selectively users are routed to the generative model. Table~\ref{tab:routing_operating_points} shows a consistent progression. Under a 1\% overall NDCG-loss budget, 2.9\% of users are routed and Novelty@10 increases by 1.2\%. Increasing the budget to 5\% routes approximately 12.0\% of users and increases novelty by 6.2\%. At the 10\% budget, approximately 20.0\% of users are routed and novelty increases by 11.1\%. The bootstrap confidence intervals remain above zero at all four operating points, and paired Wilcoxon tests indicate significant novelty improvements ($p<0.001$).

\begin{table}[t]
\centering
\small
\renewcommand{\arraystretch}{1.05}
\caption{Narrative learned-gate operating points from the probability-threshold sweep at $K=10$. Changes are relative to always serving the collaborative model. Novelty intervals are bootstrap 95\% CIs.}
\label{tab:routing_operating_points}

\begin{tabular*}{\columnwidth}
{@{\extracolsep{\fill}}cccc@{}}
\toprule
& \multicolumn{1}{c}{Routing}
& \multicolumn{2}{c}{Outcome vs.\ collaborative} \\
\cmidrule(lr){2-2}\cmidrule(l){3-4}

NDCG budget
& Routed
& $\Delta$Novelty [95\% CI]
& $\Delta$NDCG \\
\midrule

$\leq 1\%$
& $2.9\%$
& $+1.2\%$ [$1.0$, $1.5$]
& $-0.9\%$ \\

$\leq 2\%$
& $5.3\%$
& $+2.5\%$ [$2.1$, $2.9$]
& $-1.9\%$ \\

\addlinespace[1.5pt]
\textbf{$\leq 5\%$}
& \textbf{$12.0\%$}
& \textbf{$+6.2\%$ [$5.5$, $6.8$]}
& \textbf{$-4.8\%$} \\
\addlinespace[1.5pt]

$\leq 10\%$
& $20.0\%$
& $+11.1\%$ [$10.3$, $12.0$]
& $-9.6\%$ \\

\bottomrule
\end{tabular*}
\end{table}

The zero-loss condition used to construct positive training examples does not imply zero relevance loss after routing. The gate must infer the desired outcome from serving-time features and therefore makes classification errors. The routing threshold controls the consequences of these errors: conservative thresholds route only high-confidence users, whereas lower thresholds route more users and obtain additional novelty at greater overall NDCG cost. Figure~\ref{fig:routing_tradeoff}(a) visualizes this novelty--relevance trade-off.

\textbf{Learning whom to route matters.}
We next ask whether the gains arise simply from exposing more users to the higher-novelty generative model. Figure~\ref{fig:routing_tradeoff}(b) compares policies under a common overall NDCG-loss budget of at most 5\%. Importantly, routing rates are not matched: each policy selects the point on its own frontier that maximizes Novelty while satisfying the same relevance constraint. The learned gate routes 12.5\% of users and obtains a 6.5\% Novelty@10 gain. At comparable relevance cost, niche-based routing obtains 4.7\%, random routing 4.1\%, and short-history routing 2.7\%. The learned gate therefore yields a 1.8-percentage-point larger novelty gain than the strongest heuristic while spending essentially the same relevance budget. The 5\% budget is used only for comparison; deployment budgets are application-dependent.

Table~\ref{tab:routing_operating_points} uses a probability-threshold sweep, whereas Figure~\ref{fig:routing_tradeoff}(b) uses a finer coverage-ranked sweep over the same Narrative gate scores to maximize Novelty within the 5\% budget, yielding the slightly different 12.5\% routed and $+6.5\%$ Novelty point.

\textbf{The routing behavior generalizes across profile types.} Additional results in Appendix~\ref{app:additional_routing} show the same routing pattern for Centroid and TemporalNarrative. Notably, Centroid yields comparable, and at the reported points slightly stronger, novelty--relevance trade-offs than Narrative. Thus, our results support selective routing of profile-based recommenders generally, rather than an inherent frontier advantage from LLM generation itself.

Together, these results distinguish selective routing from simply replacing the collaborative model with a generative recommendation model. Generative profiles provide substantially greater exposure to novel content when used globally, but their relevance loss makes such deployment undesirable. The learned gate instead uses serving-time user characteristics to concentrate generative routing on a relatively small subset of users while exposing a threshold through which the system can choose how much relevance to exchange for additional novelty.

The study is offline and limited to one production streaming platform and the common test population across evaluated profiles. The reported operating points characterize an offline novelty--relevance frontier; online threshold calibration and robustness under distribution shift remain important future work.

\section{Conclusion}
\label{sec:conclusion}

We studied whether generative user profiles can be used selectively rather than uniformly in recommendation. While LLM-generated profiles can substantially increase novelty, deploying them globally can incur large relevance losses. Our serving-time gate instead learns when to route a user from the default collaborative sequential model to a recommendation model driven by an LLM-generated profile using only serving-time features. Varying the routing threshold exposes a controllable novelty--relevance trade-off. On real-world streaming data, the learned gate increases Novelty@10 by 6.5\% under an approximately 5\% overall NDCG-loss budget and outperforms random and simple heuristic routing at comparable relevance cost. Results with additional profile types support selective routing more generally, rather than suggesting that generative profiles universally outperform simpler semantic alternatives. Score-level fusion and unified rankers provide complementary alternatives to per-user routing; comparing these approaches under a shared candidate universe and deployment cost model is an important direction for future work.

\bibliographystyle{ACM-Reference-Format}
\bibliography{ref}

\appendix

\section{Routing Implementation and Reproducibility Details}
\label{app:routing_details}

\textbf{Representation implementation.}
Narrative and TemporalNarrative profiles are generated with Llama~3.3~70B~Instruct and encoded using SBERT \texttt{all-MiniLM-L6-v2} (384 dimensions). Centroid is constructed
by mean-pooling item embeddings produced by the same SBERT encoder.

\textbf{Routing features.}
The routing features are chosen to capture complementary aspects of a user's observable history while remaining available at serving time. History length reflects the amount of behavioral evidence available, while its logarithm distinguishes short histories without allowing very long histories to dominate the scale. Mean and median item popularity capture the user's general popularity orientation, with the median providing a more robust summary. The niche ratio measures the fraction of consumed items below the global median popularity. Finally, the norms of the collaborative and candidate profile representations provide compact model-side signals about the representations available for routing. No feature is derived from future interactions or held-out recommendation outcomes.

\textbf{Feature importance.}
Across the five out-of-fold classifiers, the collaborative representation norm is the most important feature for all three profile types, with mean importance of 0.30--0.35. The popularity characteristics of the user's interaction history are also consistently informative, while history-length features provide secondary signal. The candidate-profile norm contributes for Centroid and TemporalNarrative but is unused by the Narrative gate. These importances characterize the fitted Gradient Boosting models and should not be interpreted as causal effects.

\textbf{Serving implication.}
The routing gate is trained offline using outcome-derived binary targets. User profiles can likewise be generated and refreshed offline. At serving time, the gate uses serving-safe user features to produce a routing score, which is thresholded to select either the collaborative or profile-based recommendation model. For Narrative, the candidate-profile norm has zero importance across all five folds, so the fitted gate does not require the Narrative representation itself to make this decision.

\textbf{Gate implementation.}
A separate gate is trained for Centroid, Narrative, and TemporalNarrative using a Gradient Boosting classifier with binary log-loss and a fixed random seed of 42; otherwise, default classifier settings are used. Missing feature values are median-imputed and no feature scaling is applied. The same preprocessing and classifier
configuration are used for all three candidate recommendation models.

\textbf{Out-of-fold evaluation.}
Routing scores are obtained using five-fold stratified out-of-fold evaluation. In each fold, the gate is trained on four folds and predicts the held-out fold, so every reported routing score is produced by a
classifier that was not trained on that user's outcome-derived routing label. The resulting out-of-fold scores are then thresholded to construct the routing policies. NDCG-loss budgets are applied only when selecting
operating points after training and do not alter the classifier objective.

\section{Additional Routing Results}
\label{app:additional_routing}

Table~\ref{tab:additional_routing} reports learned-gate operating
points for Centroid and TemporalNarrative. These results complement
the detailed Narrative analysis in the main paper.

\begin{table}[h]
\centering
\small
\renewcommand{\arraystretch}{1.03}
\caption{Additional learned-gate operating points at $K=10$.
Changes are relative to always serving the collaborative model.}
\label{tab:additional_routing}

\begin{tabular*}{\columnwidth}
{@{\extracolsep{\fill}}lccc@{}}
\toprule
& \multicolumn{1}{c}{Routing}
& \multicolumn{2}{c}{Outcome vs.\ collaborative} \\
\cmidrule(lr){2-2}\cmidrule(l){3-4}
NDCG budget
& Routed
& $\Delta$Novelty
& $\Delta$NDCG \\
\midrule

\multicolumn{4}{@{}l}{\textit{Non-generative}} \\
\hspace{0.8em}\textit{Centroid} &&& \\[-2pt]

$\leq 1\%$  & $3.7\%$  & $+1.7\%$  & $-0.8\%$ \\
$\leq 2\%$  & $6.4\%$  & $+3.0\%$  & $-1.9\%$ \\
$\leq 5\%$  & $12.5\%$ & $+6.5\%$  & $-4.5\%$ \\
$\leq 10\%$ & $22.1\%$ & $+12.0\%$ & $-9.6\%$ \\

\cmidrule(lr){1-4}

\multicolumn{4}{@{}l}{\textit{Generative}} \\
\hspace{0.8em}\textit{TemporalNarrative} &&& \\[-2pt]

$\leq 1\%$  & $2.2\%$  & $+0.9\%$  & $-0.9\%$ \\
$\leq 2\%$  & $6.0\%$  & $+2.8\%$  & $-1.9\%$ \\
$\leq 5\%$  & $12.0\%$ & $+6.2\%$  & $-4.4\%$ \\
$\leq 10\%$ & $20.2\%$ & $+11.2\%$ & $-9.8\%$ \\

\bottomrule
\end{tabular*}
\end{table}

Both additional profile types exhibit the same qualitative routing pattern observed for Narrative: larger relevance budgets route more users and yield progressively greater novelty gains. Centroid achieves slightly stronger point-estimate trade-offs than Narrative at the reported budgets, while TemporalNarrative shows comparable behavior. Thus, the routing pattern is not specific to a single profile type.

\section*{Generative AI Usage Disclosure}
LLMs were used to generate the Narrative and TemporalNarrative user profiles as described in the paper. Generative AI tools were also used to assist with code and manuscript editing; all experiments, reported results, and claims were verified by the authors.

\end{document}